\documentclass[aps,prb,twocolumn,amsmath,amssymb,superscriptaddress,floatfix]{revtex4-2}

\usepackage{amsmath}
\usepackage{xcolor}
\usepackage{graphicx}
\usepackage{dcolumn}
\usepackage{bm}
\usepackage{verse}

\usepackage{here}
\usepackage{graphicx}%
\usepackage{bm}%
\usepackage{graphicx,bm}
\usepackage{pifont}
\usepackage{amsmath,amssymb}
\usepackage{color}

\newcommand{\vsc}{$\bm{\chi}_{ij}$}

\newcommand{\ssc}{$\chi_{ijk}$}
\newcommand{\TN}{$T_{\mathrm{N}}$}
\newcommand{\HT}{$H_\mathrm{t}$}
\newcommand{\HP}{$H_\mathrm{p}$}

\newcommand{\nr}{$\rho^{2f}/\rho_0$}

\begin{document}

\title{Enhanced Electrical Magnetochiral Effect\\ by Spin-hedgehog Lattice Structural Transition}

\author{A. Kitaori$^{1}$, N. Kanazawa$^{1}$, H. Ishizuka$^{2}$, T. Yokouchi$^{3}$, N. Nagaosa$^{1,4}$ and Y. Tokura$^{1,4,5}$}
\affiliation{$^{1}$ Department of Applied Physics, University of Tokyo, Tokyo 113-8656, Japan \\ 
$^{2}$ Department of Physics, Tokyo Institute of Technology, Tokyo 152-8551, Japan \\
$^{3}$ Department of Basic Science, The University of Tokyo, Meguro, Tokyo 153-8902, Japan \\
$^{4}$ RIKEN Center for Emergent Matter Science (CEMS), Wako 351-0198, Japan \\
$^{5}$ Tokyo College, University of Tokyo, Tokyo 113-8656, Japan
}

\date{\today}

\begin{abstract}
Nonreciprocal resistance, depending on both directions of current ($\bm{j}$) and magnetic-field ($\bm{H}$) or magnetization ($\bm{M}$), is generally expected to emerge in a chiral conductor, and be maximized for $\bm{j}$ $\parallel$ $\bm{H}$($\bm{M}$). This phenomenon, electrical magnetochiral effect (eMChE), is empirically known to increase with $H$ in a paramagnetic or fully ferromagnetic state on chiral lattice or to depend on fluctuation properties of a helimagnetic state. We report here the eMChE over a wide temperature range in the chiral-lattice magnet MnGe in which the spin hedgehog lattice (HL) forms with the triple spin-helix modulation vectors. The magnitude of nonreciprocal resistivity is sharply enhanced in the course of the field-induced structural transition of HL from cubic to rhombohedral form. This is attributed to the enhanced asymmetric electron scatterings by vector spin chirality in association with the large thermal fluctuations of spin hedgehogs.

\end{abstract}

\maketitle

Formation of noncollinear spin textures breaks time-reversal and/or space-inversion symmetry, leading to various responses to electromagnetic fields~\cite{nagaosa1}. These phenomena are often related to the chirality of spins, such as vector ($\bm{\chi}_{ij}= \bm{S}_i \times \bm{S}_j$) and scalar [$\chi_{ijk}= \bm{S}_i \cdot (\bm{S}_j \times \bm{S}_k)$] spin chirality. Here $\bm{S}_n$ represents the spin at the site $n$ and the spin chirality comprises adjacent multiple spins. The respective symmetry properties of \vsc\ (odd in space inversion) and \ssc\ (odd in time reversal) are transcribed into the electronic states through the spin exchange coupling and the spin-charge coupling. As consequences of the interplay between electrons and noncollinear spin textures, there emerge the electric polarization due to finite \vsc\ of cycloidal spin orders in insulators~\cite{tokura1} and geometrical/topological Hall effect due to finite \ssc\ of noncoplanar spin structures, such as two-in-two-out spin states in metallic pyrochlore oxides~\cite{taguchi1} and skyrmion-lattice states in chiral-lattice magnets ~\cite{neubauer,lee,kanazawa1,chem_rev}. 
The dynamics of noncollinear spin textures further produces rich electromagnetic properties. For instance, dynamical components of \vsc\ cause directional dichroism of light~\cite{takahashi} and the fluctuations of \ssc\ induce giant skew-scattering Hall effect~\cite{ishizuka1,yang,fujishiro}. Another remarkable example is emergent electromagnetic induction due to finite spatiotemporal spin-chirality associated with current-driven dynamics of spin helix~\cite{nagaosa_jjap,yokouchi_nature}.

Disorder or fluctuation of \vsc\ has recently been identified as a major origin of electrical magnetochiral effect (eMChE) in chiral magnets~\cite{yokouchi_comm,ishizuka2}. The eMChE is a type of nonreciprocal charge transports and its magnitude is proportional to the inner product of an applied electric current density $\bm{j}$ and an axial vector such as a magnetic field $\bm{H}$ or a magnetization $\bm{M}$~\cite{rikken,tokura2}. As a consequence of the broken inversion symmetry in a chiral material, leftward and rightward propagations of quasiparticles become asymmetric, resulting in the nonreciprocity. In MnSi and CrNb$_3$S$_6$ with chiral lattice structures, Dzyaloshinskii-Moriya (DM) interaction [$\bm{D}\cdot(\bm{S}_i \times \bm{S}_j)$] stabilizes helical spin structures~\cite{ishikawa,bak,moriya,togawa1}, which can be regarded as a sequence of \vsc\ along their propagation vectors $\bm{q}$. There have been observed the enhanced eMChE at the helimagnetic-to-paramagnetic transition or at the ferromagnetic-to-paramagnetic crossover, where fluctuations of \vsc\ proliferate~\cite{yokouchi_comm,aoki}. On the basis of the experimental observations of non-monotonous $H$-dependence of eMChE~\cite{yokouchi_comm,aoki} and the recent theoretical proposal~\cite{ishizuka2}, the chiral spin fluctuations are expected to cause asymmetric scatterings of conducting electrons, resulting in the nonreciprocal conductance. Conversely, the eMChE can be a good probe for the chiral spin fluctuation.

In this Letter, we study the eMChE in the course of the weak first-order phase transition of spin hedgehog-lattice (HL) state in the chiral magnet MnGe. At zero magnetic field, there forms the HL state constructed by the superposition of three orthogonal helical spin structures with propagation vectors ($\bm{q}_1$, $\bm{q}_2$ and $\bm{q}_3$) along the $\langle 100 \rangle$ directions~\cite{kanazawa2,tanigaki,kanazawa3}. With the application of $H$ along [111] direction, the three $\bm{q}$-vectors are discontinuously tilted towards the $H$-direction above the critical field $H_\mathrm{t}$
~\cite{kitaori}. The directional change in the basis helical structures corresponds to the structural transition from cubic [Fig.~1(a)] to rhombohedral [Fig.~1(b)] HL states, as depicted in Fig.~1(c). This magnetic transition is of a weak first-order character, possibly involving large fluctuations of hedgehogs and spin chirality~\cite{kitaori}. We observed the enhanced eMChE at the HL structural transition over a wide temperature range below the magnetic ordering temperature \TN. The close correlation between the nonreciprocal conduction profile and the fluctuating state of HL indicates the dominant role of \vsc-fluctuations, which is also supported by a simplified theoretical model dealing with a transition between helical structures with different modulation periods.

\begin{figure}[H]
\centering
\includegraphics[width =\columnwidth]{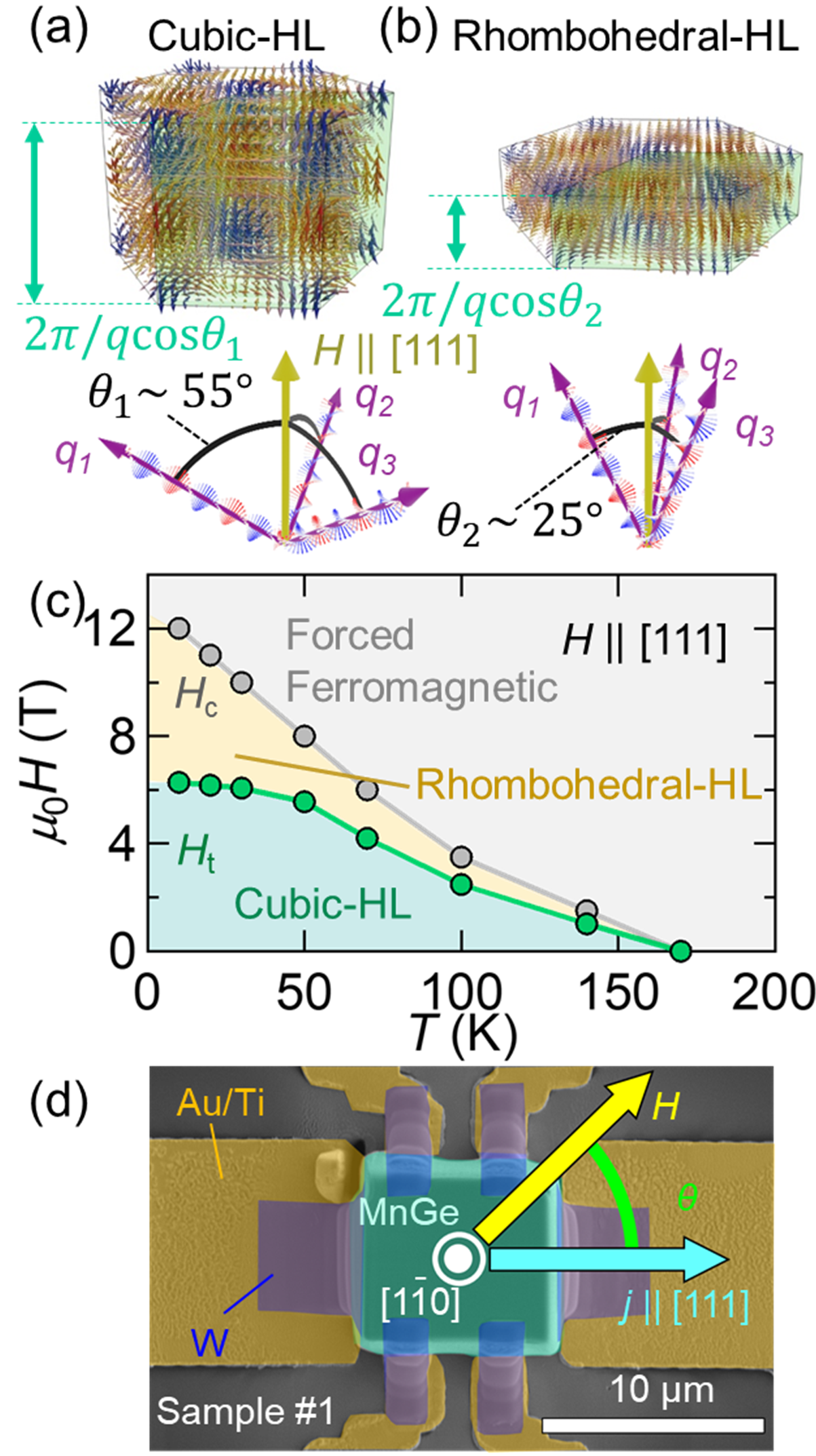}
\caption{
(a) and (b) Schematic illustrations of cubic (a) and rhombohedral (b) hedgehog lattices (HLs) as represented in terms of the hexagonal magnetic cell. Reddish and bluish arrows indicate spins with positive and negative $z$ ($\parallel [111]$) components, respectively. The spin-up (red) and spin-down (blue) clusters are alternatively stacked in the HLs, representing the three-dimensional spin-swirling texture. The hedgehogs  appear where the spin direction flip, e.g., at the midpoints between the red and blue clusters. The green boxes represent the magnetic unit cells. The height is 2$\pi/(q \cos\theta)$, where $q$ and $\theta$ are the magnitude of helical propagation vector and the angle between $q_i$ and the magnetic field. The HL-spacing is shrunk along $z$ direction upon the transformation from cubic to rhombohedral. The spin-helix bases composing the HLs are also displayed. (c) Magnetic phase diagram of MnGe under the magnetic field along [111] axis. $H_{\rm t}$ and $H_{\rm c}$ represent the critical magnetic fields for the cubic-to-rhombohedral HL transformation and the ferromagnetic transition, respectively. (d) Scanning electron microscope (SEM) image of a MnGe microdevice for the electrical transport measurements. Each component is shaded by colors for clarity: microfabricated MnGe single crystal (green), electrical pads of Au/Ti bilayer (yellow), electrical contacts of W (blue).}
\end{figure}

We estimated the nonreciprocal electrical conduction by measuring the second harmonic resistivity $\rho^{2f}$ in the microfabricated single crystals of MnGe. The single crystals were synthesized by a flux method under high pressure~\cite{kitaori}. For transport measurements, those crystals were cut into cuboids of typically $10\times 10 \times 3$-$\mathrm{\mu m}^3$ volume size by using the focused-ion-beam (FIB) technique [Fig.~1(d)]. There is little compositional variation (within 2 $\%$) among the samples, as confirmed by energy-dispersive X-ray spectroscopy. The resistivity of a chiral conductor is generally expressed in the form, $\rho=\rho_0[1+\gamma'(\bm{j}\cdot \bm{H})]$, on the basis of phenomenological arguments~\cite{rikken}. Here, $\rho_0$ is the normal resistivity independent of current density $j$, and $\gamma'$ is the coefficient of the resistivity term linear in $j$, which gives rise to the eMChE-type nonreciprocal conduction. Note that the sign of $\gamma'$ correlates with the crystal chirality. With application of an ac input current density $j = j_0 \sin(2\pi f t)$, the output electric-field reads $E=\rho j=\rho_0 j_0 \sin(2\pi f t) + \frac{1}{2}\gamma'\rho_0Hj_0^2\cos\theta \left[1+\sin\left(4\pi f t -\frac{\pi}{2}\right) \right]$, where $\theta$ is the angle between $\bm{j}$ and $\bm{H}$. The nonreciprocal resistivity is thus estimated by measuring the second harmonic voltage with $-\pi/2$ shift~\cite{pop}: $\rho^{2f}=\frac{\mathrm{Im}[E^{2f}]}{j_0}=-\frac{1}{2}\gamma'\rho_0Hj_0 \cos\theta$. We applied the current and mangetic field along MnGe[111] direction [Fig.~1(d)]. Current density and frequency are typically $j_0= 3.5 \times 10^8$ A/m$^2$ and $f=37$ Hz.

Figures~2(a)-(e) show $H$-dependence of $\rho^{2f}$ normalized by $\rho_0$ at $\theta = 0$. Below $T=T_{\mathrm{N}}$, the profile of \nr\ comprises two components: a peak structure at $H=H_{\mathrm{p}}$ and an $H$-linear term. The peak width becomes broader with decreasing $T$, while no significant change is observed in the peak magnitude. The magnitude of the $H$-linear component increases especially below $T\sim 20$ K and its sign reversal occurs below $T\sim 55$ K. Since no kink structure is discerned at $H=H_\mathrm{c}$, where $M$ saturates upon the transition to the forced ferromagnetic state, there seems little nonreciprocal resistivity proportional to $M$ (also see Fig.~S1 ~\cite{supple}).
The $j$-dependence of nonreciprocal resistivity is shown in Fig.~2(f). Both the two components, which are respectively estimated as \nr\ at $H=H_{\mathrm{p}}$ and 14 T, follow the $j$-linear dependence as expected in the case of nonreciprocal resistivity due to eMChE. Here we note that no clear nonreciprocal signal is observed around $T_{\rm N}$, unlike the case of MnSi ~\cite{yokouchi_comm}. This may be because \vsc\ -fluctuations with positive and negative signs may be simultaneously excited and the net \vsc\ may become negligibly small above the high $T_{\rm N}$ ($\sim 170$ K) in MnGe, while the spin fluctuations exhibit the homochiral nature just above $T_{\rm N}$ ($\sim 30$ K) in MnSi ~\cite{shirane,grigoriev1,pappas, grigoriev2}.

\begin{figure*}
\includegraphics[width=17cm]{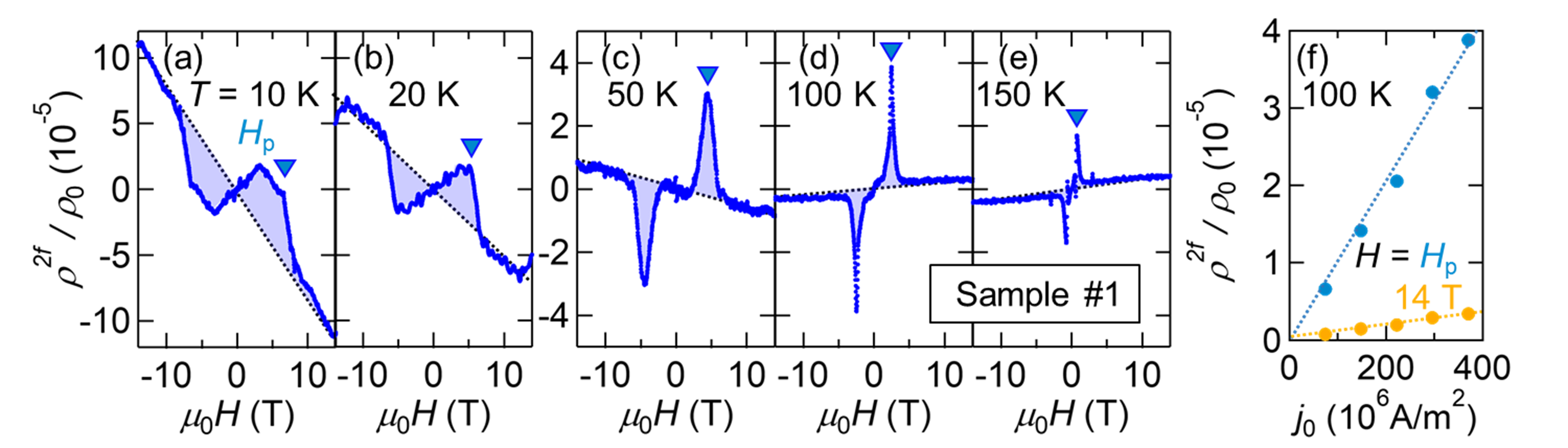}
\caption{
(a)-(e) Magnetic-field $H$ dependence of normalized nonreciprocal resistivity \nr\ at various temperatures under $\bm{H} \parallel \bm{j}$ with the current density of $j_0= 3.5 \times 10^8$ A/m$^2$. The blue triangles indicate the magnetic fields \HP\ at the tops of positive peak structures. The black dashed lines are the guides for the $H$-linear components of \nr. (f) Current-density ($j_0$) dependence of \nr\ at $H=H_{\mathrm{p}}$ and 14 T. The data were measured at $T=100$ K. The color dashed lines indicate the $j_0$-liner fittings.}
\end{figure*}

We checked the nonreciprocal conduction properties of four different samples (see Fig.~S2 ~\cite{supple}). Although the overall $H$-dependence of \nr\ is essentially the same among the different devices, its magnitude varies: the peak value of \nr\ at $H=H_{\mathrm{p}}$ falls within the relatively narrow range between 3--$5.5\times 10^{-5}$; the $H$-linear component of \nr\ at $H=14$ T shows the considerable variation between 0.5--$8.5\times 10^{-5}$. Such sample-dependence implies that electron scatterings are involved crucially in the nonreciprocal conduction. Also, the dominant scattering mechanisms causing the two nonreciprocal components would be different because their sample dependence is different. In particular, the peak structure is less sensitive to sample differences. This suggests that the peak structure stems inherently from scatterings by magnetic structure or its fluctuation, rather than those by crystal imperfections.

We show in Fig.~3 the magnetic-field direction dependence of \nr\ at $T=50$ K to confirm that the nonreciprocal resistivity originates from eMChE. With changing from $\theta=0^{\circ}$ to $\theta=180^{\circ}$, \nr\ shrinks in magnitude, subsequently reverses its sign above $\theta=90^{\circ}$, and then returns to the original magnitude with the opposite sign [Fig.~3(a)]. To separably scrutinize the two components of \nr, we extract \nr\ at $H=4$ T ($\sim H_\mathrm{p}$) and $H=14$ T as functions of $\theta$ in Figs. 3(b) and (c). Both the \nr\ vs. $\theta$ curves obey the $\cos\theta$-variation, which are consistent with the eMChE-type nonreciprocal conduction in proportion to $\bm{j}\cdot\bm{H}$. The opposite signs of $\theta$-dependence [Figs. 3(b) and (c)] again highlight that the two nonreciprocal components originate from different scattering mechanisms. Here we note that a small deviation of \nr\ from the $\cos\theta$ curve is observed at $H=4$ T [Fig.~3(c)]. This suggests that the eMChE may be derived from the complex \vsc-distribution, which cannot completely follow $\bm{H}$-direction in the course of the cubic-to-rhombohedral HL transformation. The HL structure is possibly dependent on the $H$-direction and is deformed with rotating $H$.

To examine the relation between the nonreciprocal resistivity and the spin state, we show a contour map of \nr\ overlaid on the $T$-$H$ phase diagram [Fig.~4(a)]. The positive \nr\ shows up around the cubic-rhombohedral HL phase boundary and its peak (\HP) correlates with the HL phase transition field \HT\ over the wide $T$-range below \TN. The coincidence of the positive \nr\ and the HL transformation can be assigned to the asymmetric scattering of electrons by the proliferated \vsc-fluctuations. Meanwhile, the $H$-linear component of \nr\ becomes pronounced at low $T$, being irrelevant to the spin state. A possible origin of the sharp increase with lowering $T$ is an emergence of high-mobility carriers like the Weyl fermions, which have been identified in the isostructural crystals~\cite{chang,tang,rao,sanchez}. In any case, the two components would be rooted in the different scattering mechanisms as also evident from the large difference in their $T$-dependence [Fig.~4(b)].

\begin{figure}[H]
\centering
\includegraphics[width =\columnwidth]{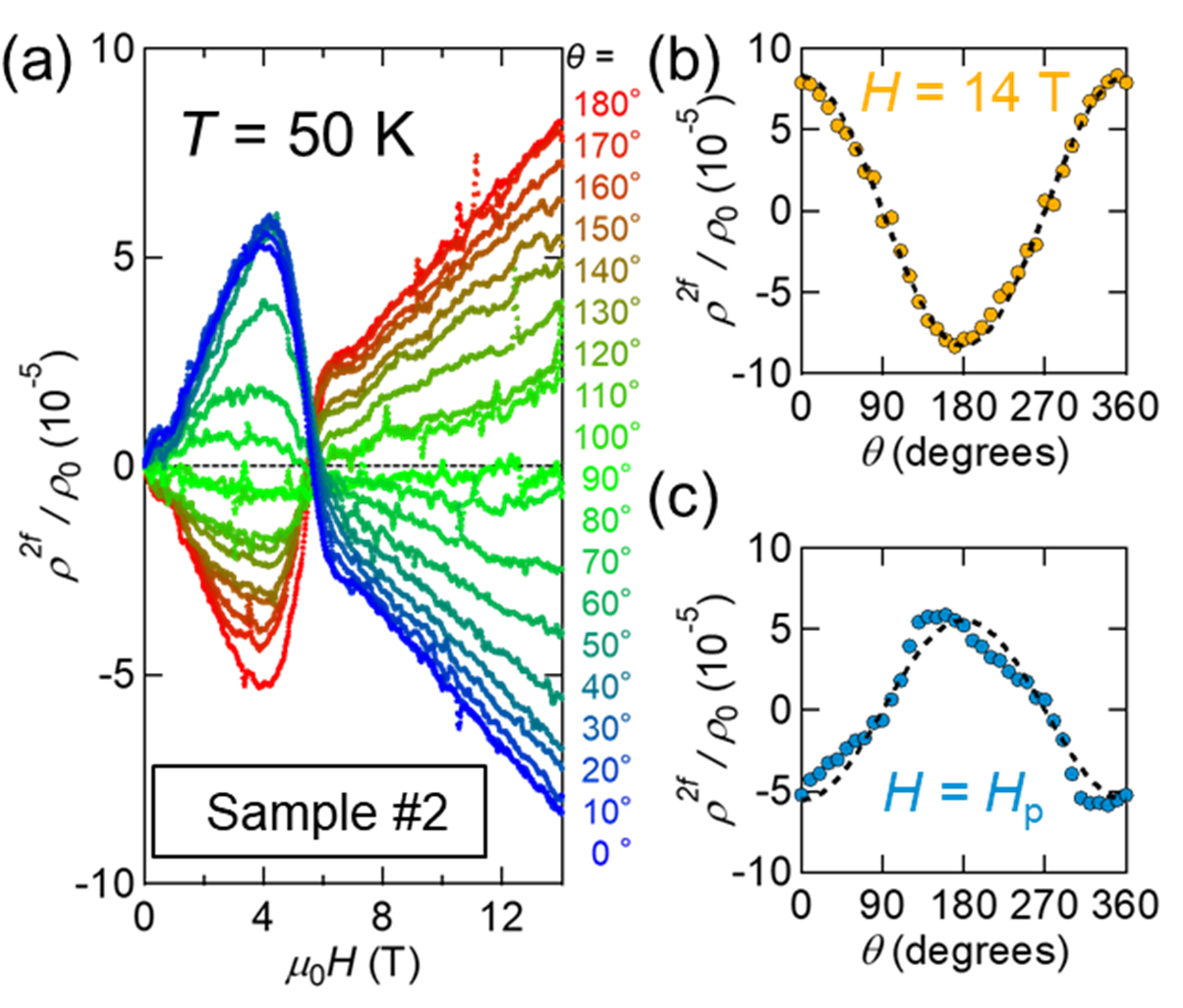}
\caption{
(a) Magnetic-field $H$ dependence of normalized nonreciprocal resistivity \nr\ under various $H$-directions $\theta=0^{\circ}$-$180^{\circ}$, theta being the angle between $\bm{H}$ and $\bm{j}$ (both in-plane). The current density is set at $j_0= 3.5 \times 10^8$ A/m$^2$. (b) and (c) Angular $\theta$ dependence of \nr\ at $H=14$ T (b) and $H=4$ T (c), which respectively exemplify the $H$-linear component and the peak component at the HL transformation. The dashed lines are the fitting curves of $\cos\theta$-variation. The data in panels (a)-(c) were measured at $T=50$ K.}
\end{figure}

Below, we verify the scenario that the \vsc-fluctuation causes the peak structure of \nr\ by employing a spin model that approximates the HL transformation. As is the case for the conventional eMChE which is maximized when $j$ flows along the screw axis of the crystal, the component of \vsc\ parallel to $\bm{j}$ contributes to the nonreciprocal resistance. Therefore, the number of helical turns along $\bm{j}$ ($\parallel\bm{H}$) is the essential factor. 
Here we replicate the shrinkage of HL-spacing along $H$ [see Figs. 1(a) and (b)] by considering the contraction of the modulation period of the helical spin structure. The spin model which we used is:
\begin{eqnarray*}
\mathcal{H}= &-&J_0 \sum_{\langle i,j\rangle \in xy\mathrm{\mathchar`-plane}} \bm{S}_i \cdot \bm{S}_j 
-J_1 \sum_{\langle i,j\rangle \in \mathrm{n.n.}} \bm{S}_i \cdot \bm{S}_j\\
&-&J_2 \sum_{\langle i,j\rangle \in \mathrm{2n.n.}} \bm{S}_i \cdot \bm{S}_j
-J_3 \sum_{\langle i,j\rangle \in \mathrm{3n.n.}} \bm{S}_i \cdot \bm{S}_j \\
&-&J_4 \sum_{\langle i,j\rangle \in \mathrm{4n.n.}} \bm{S}_i \cdot \bm{S}_j
-D \sum_{\langle i,j\rangle \in \mathrm{n.n.}} \hat{\bm{z}}\cdot (\bm{S}_i \times \bm{S}_j),
\end{eqnarray*}
where ${J}_{0}=\frac{1}{2}$, ${J}_{1}=1$, ${J}_{2}=\frac{1}{\sqrt{2}}+\frac{D}{2\sqrt{2}}$, ${J}_{3}=\frac{1}{3}$, ${J}_{4}=-\frac{1}{3\sqrt{3}}+D(\frac{3}{4\sqrt{2}}-\frac{1}{2})+\delta$ and $D=0.1$ (see Supplemental Material for details ~\cite{supple}). The first term of $\mathcal{H}$ represents the ferromagnetic interaction in the $xy$-plane. The second to fifth terms are exchange interactions between $n$-th nearest neighbor sites along $z$-direction ($n=$1-4). The last term represent the DM interaction, which favors a specific spin helicity. In this model, the ground state changes between eightfold-period ($\delta < 0$) and fourfold-period ($\delta > 0$) helical structures by changing $\delta$ in $J_4$, which controls the frustration among the exchange interactions along $z$-direction $J_{1-4}$. The corresponding helical propagation vectors along $z$ are $q=\pi/4$ ($\delta<0$) and $q=\pi/2$ ($\delta>0$), respectively (See Fig.~S3 ~\cite{supple}). The first-order phase transition between the two helical states mimics that of the phase transition in MnGe under the magnetic field.\\

Here we focus on the fluctuation contribution to the average vector spin chirality $\Delta\chi$. We note that there are two contributions to the average vector chirality in the ordered phase : one from the magnetic order and another from the magnetic fluctuation (or spin waves with nonzero chirality). Among the two contributions, only the fluctuation contribution produces nonreciprocal current in the scattering scenario. We estimate the fluctuation contribution by $\Delta\chi\equiv\chi-\chi^{\mathrm{order}}$ where $\chi\equiv\frac1{N}\sum_{ij}\bm z\cdot(\bm S_i\times\bm S_j)$ is the thermal average of the vector chirality, and $\chi^{\mathrm{order}}$ is the mean-field contribution defined by $\chi^{\mathrm{order}}=2 S_q \sin(q)$ ($S_q$ is the spin structure factor and $q$ is the wave number). The development of $\Delta\chi$ upon the transition of helical period is presented in Fig.~4(c). With increasing $\delta$ from negative to positive, $\Delta\chi$ gradually increases in the $q=\pi/4$ phase, and shows rapid reduction on entering the $q=\pi/2$ phase. This profile resembles the $H$-dependence of nonreciprocal resistivity as we can see \nr\ as functions of $H$ scaled by \HT\ in Fig.~4(d): The peak structure asymmetrically stretches to the cubic-HL phase. The broadening of peak structure especially at low $T$ may originate from the enhanced magnetic anisotropy, which more strongly pins the hedgehogs and promotes nonlinear distortion of the magnetic structure, bringing about disorder in \vsc.

\begin{figure}[H]
\centering
\includegraphics[width =\columnwidth]{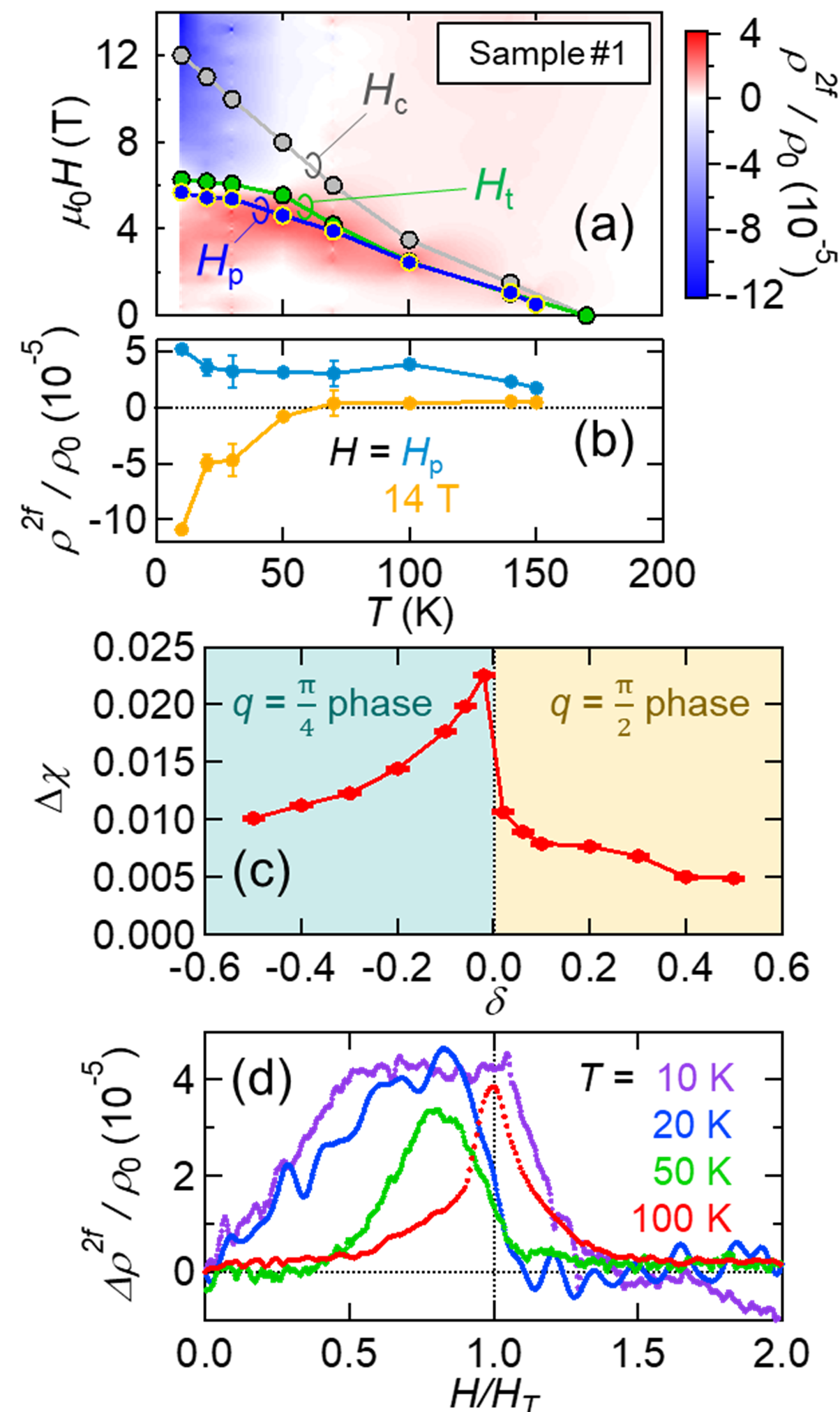}
\caption{
(a) Contour map of normalized nonreciprocal resistivity \nr\ along with characteristic fields \HP\ (for \nr\ peak), \HT\ (for the cubic-rhombohedral HL structural transition) and $H_{\mathrm{c}}$ (for the field-aligned ferromagnetic transition). (b) Temperature dependence of \nr\ at $H=H_{\mathrm{p}}$ and 14 T. (c) Simulated results for the fluctuation component of vector spin chirality $\Delta\chi$ around the transition of helical period between $q=\pi/4$ and $q=\pi/2$ in a spin model to replicate the HL transformation. $\delta$ is a parameter in $J_4$, which modulates the frustration among the exchange interactions $J_{1-4}$. Simulations are performed at $T = 0.2 J_1$. (d) The peak structures of \nr\ as functions of $H$ scaled by \HT\ at various temperatures.
}
\end{figure}

In summary, we have observed the enhanced nonreciprocal magnetotransport due to the electrical magnetochiral effect around the cubic-to-rhombohedral transformation of the spin hedgehog lattice (HL) in MnGe. We identified that the asymmetric scatterings of conduction electrons by the fluctuation of vector spin chirality causes the prominent nonreciprocal conduction. By focusing on the weak first-order transition boundary of the HL transformation, which traverses the entire ordered phase diagram on the temperature-magnetic field plane, we could exploit the proliferated spin-chirality fluctuations to realize the nonreciprocal conduction in a wide $T$-range. Although the simplified theoretical model could reproduce the qualitative behavior of chiral fluctuations at the first-order transition of helical period, more realistic models will be desired for revealing a possible role of topological nature of spin hedgehogs and for elucidating the nonreciprocal resistivity with the vector-spin-chirality fluctuations in a more quantitative way. In particular, the magnetic excitations of HL states may show unique low-energy spectral features, reflecting the dynamics of topological point defects. This would be clarified by recent theories for three-dimensional topological spin texture ~\cite{grytsiuk,tapia,okumura,shimizu} or can be directly observed by inelastic neutron scattering techniques ~\cite{martin}. Our findings highlight the large fluctuations of spin hedgehogs as a source of nonreciprocal conduction, which may enable the magnetically-controllable rectification of electron flows. Systematic investigations of nonreciprocal conduction in various topological magnets remain as future challenges.

\section*{Acknowledgments}
We thank Y. Fujishiro and F. Kagawa for fruitful discussions. This work was supported by JSPS KAKENHI (Grants No. JP20H01859, No. JP20H01867 and No. JP20H05155) and JST CREST (Grant No. JPMJCR16F1 and No. JPMJCR1874).

\newpage
\onecolumngrid
\setcounter{equation}{0}
\setcounter{figure}{0}
\setcounter{table}{0}
\setcounter{page}{1}
\makeatletter
\renewcommand{\theequation}{S\arabic{equation}}
\renewcommand{\thefigure}{S\arabic{figure}}
\renewcommand{\bibnumfmt}[1]{[S#1]}

\renewcommand{\citenumfont}[1]{S#1}

\maketitle

\section*{Supplemental Material}

\section*{Resistivity and magnetization properties in MnGe}
We show in Fig. S1 magnetic-field $H$ dependence of magnetization $M$ and resistivity term $\rho_0$ constant with respect to current density $j$, in comparison with that of the normalized nonreciprocal resistivity \nr. We note that we could not measure $M$ of the MnGe single crystal because of the small crystal size. Here we instead show $M$ of a thick epitaxial film of MnGe with comparable thickness ($\sim$3 $\mu$m) ~\cite{kitaori}. At the ferromagnetic transition ($H=H_\mathrm{c}$), $M$-$H$ curves bend in accord with the saturation of $M$. The $\rho_0$ exhibits bump structures at the HL transformation ($H=H_\mathrm{t}$) or at the $H$ just below $H_\mathrm{c}$, which are attributed to the large spin fluctuations upon the HL transformation~\cite{kitaori} and the pair annihilation of hedgehogs~\cite{kanazawa2}, respectively. As discussed in the main text, the positive peak structure is observed in \nr\ at $H=H_\mathrm{t}$, while no kink structure is discerned in \nr\ upon the ferromagnetic transition $H=H_\mathrm{c}$. This indicates little contribution from the eMChE in proportion to $M$ or asymmetric scatterings by reduced \vsc-fluctuations upon the pair annihilation.

\begin{figure*}[h]
\begin{center}
\includegraphics*[width=13cm]{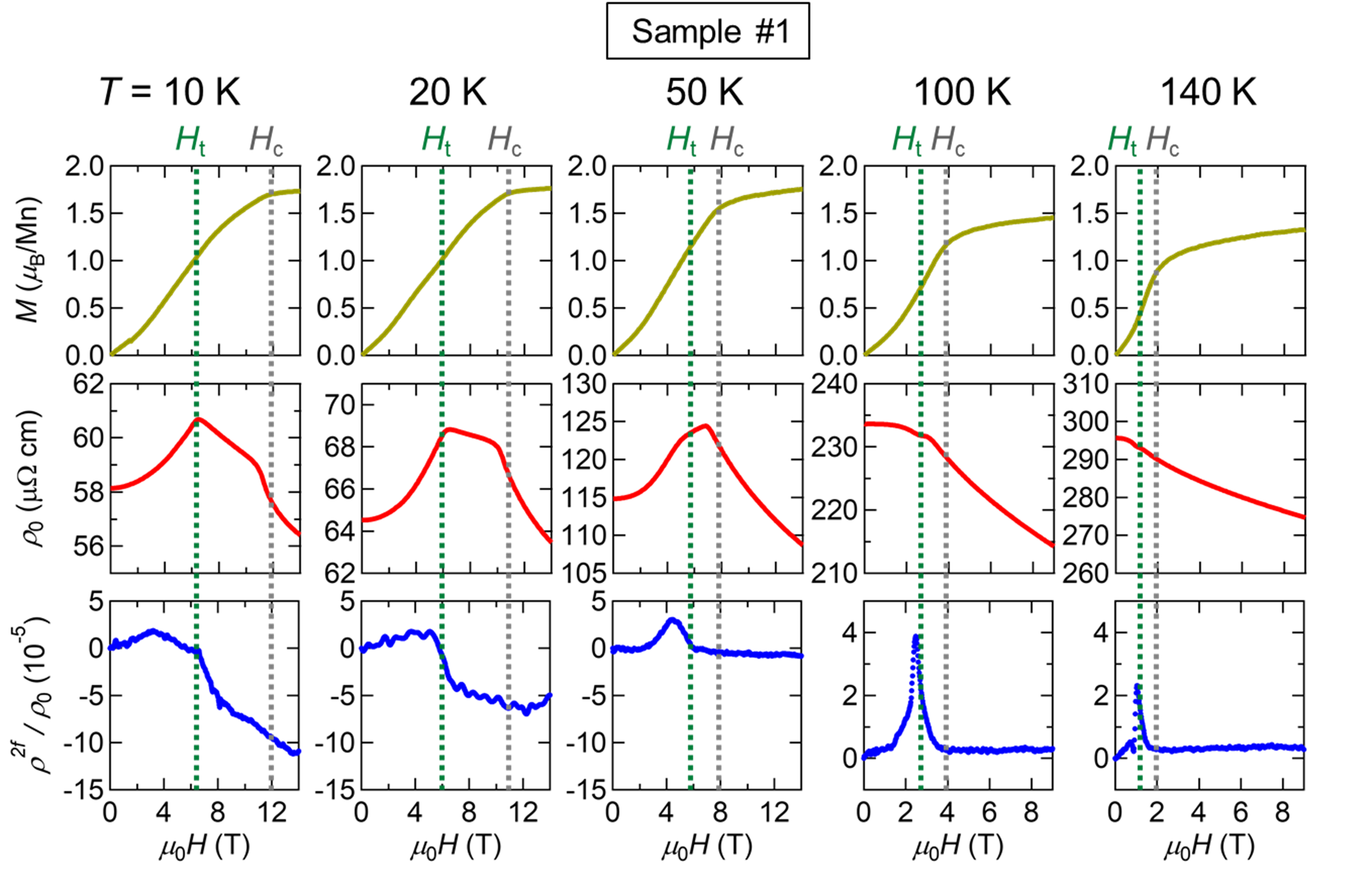}
\caption{
Magnetic-field $H$ dependence of magnetization $M$ (upper panels), resistivity $\rho_0$ (middle panels) and normalized nonreciprocal resistivity \nr\ (lower panels) at various temperatures $T$. Vertical dashed lines indicate the critical fields $H_\mathrm{t}$ for the hedgehog-lattice transformation and $H_\mathrm{c}$ for the ferromagnetic transition.}
\end{center}
\end{figure*}

\section*{Sample dependence of nonreciprocal resistivity}
We measured nonreciprocal resistivity in four different samples to check the reproducibility. Figure S2 shows $H$-dependence of \nr\ at $T=50$ K. Despite the difference in signal magnitude, the profiles of \nr\ exhibit the similar $H$-dependence among all the samples, {\it i.e.}, the peak structure at the hedgehog-lattice (HL) transformation and the $H$-linear component. Here we note that the sign of \nr\ in sample \#4 is opposite to the others'. The sign reversal can be attributed perhaps to the opposite crystalline chirality between sample \#4 and the others.

\begin{figure*}[htbp]
\begin{center}
\includegraphics*[width=12cm]{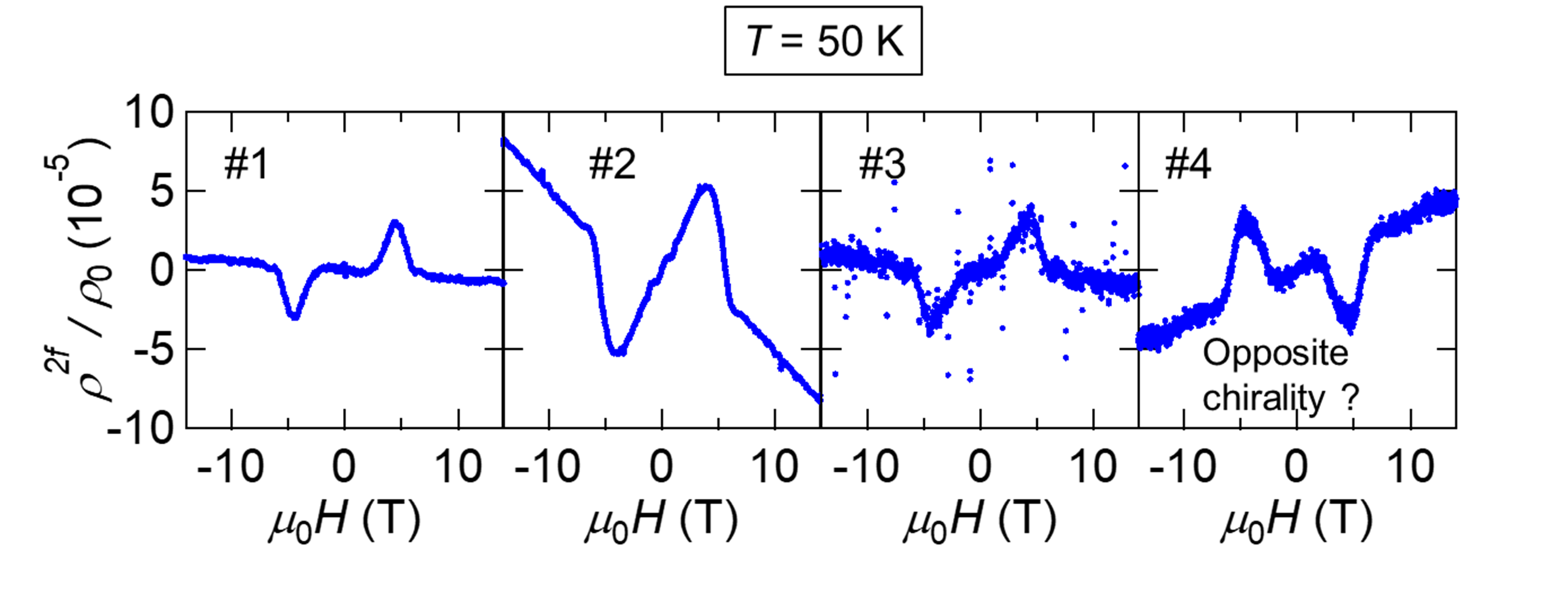}
\caption{Magnetic-field $H$ dependence of normalized nonreciprocal resistivity \nr\ in four different samples. The opposite sign of \nr\ between sample \#4 and the others indicates that sample \#4 has the opposite crystalline chirality to the other samples.}
\end{center}
\end{figure*}

\section*{Monte Carlo method}
The theoretical calculation of vector chirality at a phase boundary is based on a Monte Carlo (MC) simulation of the model in the main text. We used the classical Monte Carlo method with the heat-bath update method to calculate the chirality at the phase boundary. Note here that the helical period is devised to be determined by the frustration in the $J_{1-4}$, not by the $J_0$/$D$ ratio, which is the case for MnGe with short helical periods of a few nm scale in contrast with the FeGe case~\cite{shibata}. The data shown in Fig. 4 (c) is for $N$=16×16×32 site system at $T/{J}_{1}$=0.2. We averaged over 100000 MC steps after discarding the first 10000 steps to obtain the thermal average. These samplings were divided into five bins to estimate the statistical error. Figure S3 shows the resultant magnetic phase diagram of the spin model. 

\begin{figure*}[h]
\begin{center}
\includegraphics*[width=6cm]{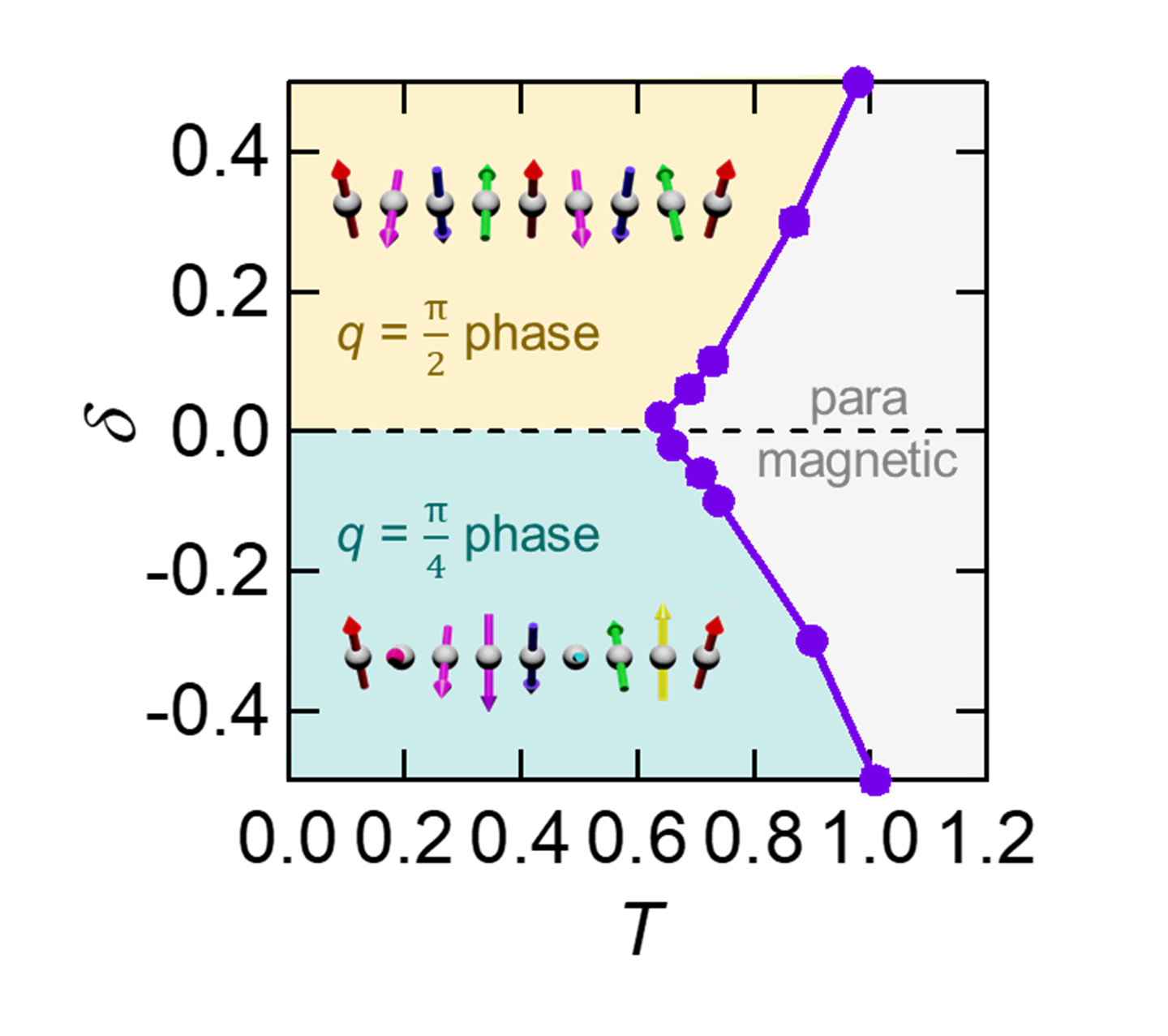}
\caption{
Magnetic phase diagram of the employed spin model. (See main text for details.) Schematic illustrations for the single-$q$ states are presented in the inset.}
\end{center}
\end{figure*}

\end{document}